\documentclass{article}

\usepackage{amssymb}
\usepackage[usenames,dvipsnames]{xcolor}
\usepackage{subfig}
\usepackage{graphicx}
\usepackage{amsmath}
\usepackage{amssymb}
\usepackage{mathbbol}

\usepackage{url}

\def\BMP{ \otimes } % Boolean matrix product
\def\sep{} %separates two indexes for instance: $x_{i \sep j}$

\def\allInd{\ast }

\def\lsetset{\mathbb{L}}
\def\eeins{\mathbb 1}

\begin{document}

\author{
Mario Frank, Ben Dong, Adrienne Porter Felt, Dawn Song
\\
{
\footnotesize
University of California, Berkeley
}
}

\title{Mining Permission Request Patterns from Android and Facebook Applications (extended author version)}

\maketitle

\begin{abstract}
Android and Facebook provide third-party applications with access to users'
private data and the ability to perform potentially sensitive operations (e.g.,
post to a user's wall or place phone calls). As a security measure, these
platforms restrict applications' privileges with permission systems: users must
approve the permissions requested by applications before the applications can
make privacy- or security-relevant API calls. However, recent studies have shown
that users often do not understand permission requests and lack a notion of typicality of requests. As a first step towards
simplifying permission systems, we cluster a corpus
of 188,389 Android applications and 27,029 Facebook applications to find
patterns in permission requests. Using a method for Boolean matrix factorization for finding overlapping clusters, we find that Facebook permission requests
follow a clear structure that exhibits high stability when fitted with only five clusters, whereas
Android applications demonstrate more complex permission requests. We also find
that low-reputation applications often deviate from the permission request patterns that we identified for high-reputation applications suggesting that permission request patterns are indicative for user satisfaction or application quality.
\end{abstract}

\section{Introduction}

Open development platforms like Android and the Facebook Platform have resulted
in the availability of hundreds of thousands of third-party applications that
end users can install with only a few clicks. Consequently, end users are faced
with a large and potentially bewildering number of choices when looking for
applications. Users' installation decisions have privacy and security
ramifications: Android applications can access device hardware and data, and
Facebook applications can access users' profile information and social networks.
As such, it is important to help users select applications that operate as the
user intends.

Android and Facebook use permission systems to control the privileges of
applications. Applications can only access privacy- and security-relevant
resources if the user approves an appropriate permission request. For example, an Android application can only send text messages if
it has the \verb|SEND_SMS| permission; during installation, the user will see a
warning that the application can ``Send SMS messages'' if the installation is
completed. These permission systems are intended to help users avoid privacy- or
security-invasive applications. Unfortunately, user research has demonstrated
that many users do not pay attention to or understand the permission
warnings~\cite{Felt:EECS-2012-26,King2011}. A major problem here is that users do not know what permission combinations are typical for applications.

Our work is a first step in the direction of simplifying permission systems by
means of statistical methods. We propose to identify common patterns in
permission requests so that applications that do not fit the predominant
patterns can be flagged for additional user scrutiny. Towards this goal, we
apply a probabilistic method for Boolean matrix factorization to the permission
requests of Android and Facebook applications. We find that while applications
with good reputations (i.e., many ratings and a high average score) typically
correspond well to a set of permission request patterns, applications with poor
reputations (i.e., less than 10 ratings) often deviate from those patterns.

The primary contribution of this paper is the first analysis of permission
request patterns with a statistically sound model. Our technique captures the
concept of identifying ``unusual'' permission requests. Our evaluation
demonstrates that our technique is highly generalizable, meaning that 
the found clustering is stable over different random subsets of the data. %we have likely discovered the true statistical structure of the data.
 We find that
permission request patterns can indicate user satisfaction or application
quality.

\section{Background and Related Work}

Android and the Facebook Platform support extensive third-party application
markets. They use permission systems to limit applications' access to users'
private information and resources.

\subsection{Android}

The Android Market is the official (and primary) store for Android
applications. The Market provides users with average user ratings, user
reviews, descriptions, screenshots, and permissions to help them select applications.
Android applications can access phone hardware (e.g., the
microphone) and private user information (e.g., call history) via Android's API. Permissions
restrict applications' ability to use the API. For example, an application can
only take a photograph if it has the \verb|CAMERA| permission. Developers select
the permissions that their applications need to function, and these permission
requests are shown to users during the installation process.
The user must approve all of an application's
permissions in order to install the application.

Several studies have examined Android applications' use of permissions. Barrera
et al.\ surveyed the $1,100$ most popular applications and found that
applications primarily request a small number of permissions, leaving most other
permissions unused~\cite{somAndroid}. They used self-organizing maps (a
dimensionality reduction technique) to visualize the relationship between
application categories and permission requests; based on this analysis, they
concluded that categories and permissions are not strongly related. Their focus
was on visualization and their findings are not applicable to identifying
unusual permission request patterns; they relied on the minimization of a
Euclidian cost function to find a low dimensional visualization of the data,
whereas we use a generative probabilistic model to learn request patterns. Felt
et al.\ and Chia et al.\ surveyed Android applications and identified the
most-requested permissions~\cite{chia12,felt-webapps}. Chia et al.\ also found
several correlations between the number of permissions and other factors: a weak
positive correlation with the number of installs, a weak positive correlation
with the average rating, a positive correlation with the availability of a
developer website, and a negative correlation with the number of applications
published by the same developer. We expand on these past analyses, and our analysis of
permission requests is by far the largest study to date.

Other research has focused on using machine learning techniques to identify
malware. Sanz et al.\ applied several types of classifiers to the permissions,
ratings, and static strings of $820$ applications to see if they could predict
application categories, using the category scenario as a stand-in for malware
detection~\cite{Sanz2012androidappclass}. Shabtai et al.\ similarly built a
classifier for Android games and tools, as a proxy for malware
detection~\cite{Shabtai:2010:ASC:1931473.1932178}. Zhou et al.\ found real
malware in the wild with DroidRanger, a malware detection system that uses
permissions as one input~\cite{droidranger}. Although our techniques are
similar, our goal is to understand the difference between high-reputation and
low-reputation applications rather than to identify malware. Applications may be
of low quality or act in undesirable ways (i.e., be risky) without being
malware. Additionally, our approach only relies on permission requests; unlike
these past approaches, we do not statically analyze applications to extract
features, which makes our technique applicable to platforms where code is not
available (such as the Facebook platform).

Enck et al.\ built a tool that warns users about applications that request
blacklisted sets of permissions~\cite{kirin}. They manually selected the
blacklisted patterns to represent dangerous sets of permissions. In contrast, we
advocate a statistical whitelisting approach: we propose to warn users about
applications that do not match the permission request patterns expressed by
high-reputation applications. These two approaches could be complementary; human
review of the statistically-generated patterns could potentially improve them.

\subsection{Facebook}

The Facebook Platform supports third-party integration with Facebook. Facebook
lists applications in an ``Apps and Games'' market alongside information about
the applications, including the numbers of installs, the average ratings, and
the names of friends who use the same applications. Through the Facebook
Platform, applications can read users' profile information, post to users' news
feeds, read and send messages, control users' advertising preferences, etc.
Access to these resources is limited by a permission system, and developers must
request the appropriate permissions for their applications to function. Applications can request permissions at any time, but most permission requests are displayed during installation as a condition of installation. Chia
et al. surveyed Facebook applications and found that their permission usage is
similar to Android applications: a small number of permissions are heavily used,
and popular applications request more permissions~\cite{chia12}.

\section{Application Data Set}

\subsection{Data Collection} \label{sec_datacollect}

\subsubsection{Android} We collected information about 188,389 Android
applications from the official Android Market in November 2011\footnote{The data is available at \url{http://www.mariofrank.net/andrApps/index.html}}. This data set
encompasses approximately $59\%$ of the Android Market, which contained 319,161
active applications as of October 2011~\cite{marketsize}. To build our data set,
we crawled and screen-scraped the web version of the Android Market. Each
application has its own description page on the Market website, but the Market
does not provide an index of all of its applications. To find applications'
description pages, we first crawled the lists of ``top free'' and ``top paid''
applications. These lists yielded links to 32,106 unique application pages.
Next, we fed 1,000 randomly-selected dictionary words and all possible
two-letter permutations (e.g., ``ac'') into the Market's search engine. The
search result listings provided us with links to an additional $156,283$ unique
applications. Once we located applications' description pages, we parsed their
HTML to extract applications' names, categories, average rating score, numbers
of ratings, numbers of downloads, prices, and permissions.

\subsubsection{Facebook} Chia et al. provided us with a set of 27,029 Facebook
applications~\cite{chia12}. They collected these applications by crawling
SocialBakers~\cite{socialbakers}, a site that aggregates statistics about
Facebook applications. After following SocialBakers's links to applications,
they screen-scraped any permission prompts that appeared. They also collected
the average ratings and number of ratings for each application. One limitation
of this data set is that it only includes the permission requests that are shown
to users as a condition of installation; they did not attempt to explore the
functionality of the applications to collect secondary permission requests that
might occur later. As such, our analysis only incorporates the permission
requests that are shown to users as part of the installation flow.

\subsection{Global Statistics of the Dataset} \label{globalStats}

As an overview, we provide global statistics of the application datasets. We
investigate overall application features, such as the price, ratings, and most
popular permissions. The characteristics of these features play a role in our
analysis of permission request patterns (Section~\ref{sec_Exps}).

\paragraph{Permission Requests.} Table~\ref{tab_globalperms} lists the 15 most
frequently requested Android permissions, and Table~\ref{tab_FB} depicts the
15 most frequently requested Facebook permissions. As these indicate, a small
number of permissions are widely requested, but most permissions are
infrequently requested.

\begin{table}
\centering
\caption{15 most frequently requested Facebook permissions out of 62 permissions. \label{tab_FB}\vspace{-0.5em}
} 
\begin{tabular}{| l | l | }
\hline
requested & permission name\\
\hline
67.35\% & basic\\
23.12\% & publish\_stream\\
13.93\% & email\\
3.38\% & user\_birthday\\
2.47\% & offline\_access\\
2.12\% & user\_photos\\
1.79\% & publish\_actions\\
1.52\% & user\_location\\
1.32\% & read\_stream\\
1.16\% & user\_likes\\
1.05\% & user\_about\_me\\
0.95\% & friends\_photos\\
0.75\% & user\_hometown\\
0.69\% & user\_videos\\
0.6\% & friends\_birthday\\
\hline
\end{tabular}
\end{table}

{
\begin{table}
\centering
\caption{15 most frequently requested Android permissions out of 173 total permissions. \label{tab_globalperms}}\vspace{-0.5em}
\begin{tabular}{| l | l | }
\hline
requested & permission name\\
\hline
69.76\% & Network communication : full Internet access \\
43.24\% & Network communication : view network state \\
30.26\% & Storage : modify/delete USB storage \& SD card contents \\
26.47\% & Phone calls : read phone state and identity \\
18.34\% & Your location : fine (GPS) location \\
16.89\% & Your location : coarse (network-based) location \\
16.16\% & Hardware controls : control vibrator \\
15.01\% & System tools : prevent device from sleeping \\
8.22\% & Network communication : view Wi-Fi state \\
8.11\% & System tools : automatically start at boot \\
6.71\% & Services that cost money: directly call phone numbers\\
6.27\% & Your personal information : read contact data \\
5.59\% & Hardware controls : take pictures and videos \\
4.61\% & System tools : set wallpaper \\
3.9\% & System tools : retrieve running applications \\
\hline
\end{tabular}
\end{table}
}

\begin{figure*}[htb]
\centering
\subfloat[Cumulative Price Distribution]{\label{fig_prices}\includegraphics[width=0.44\textwidth]{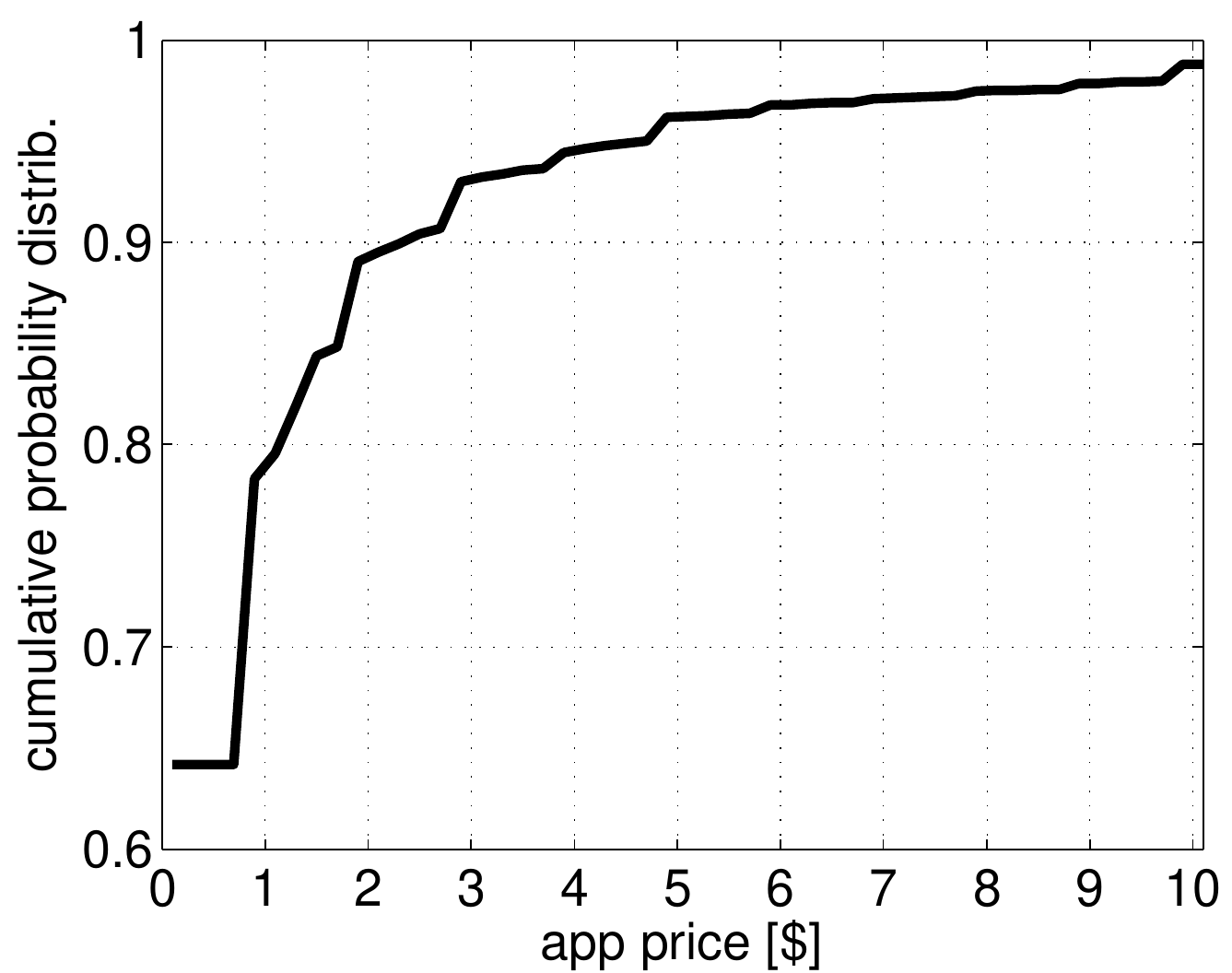}
  }
 \ \ \ \ \ %// %~
\subfloat[Rating Distribution ]  {  \label{fig_ratings} \includegraphics[width=0.47\textwidth]{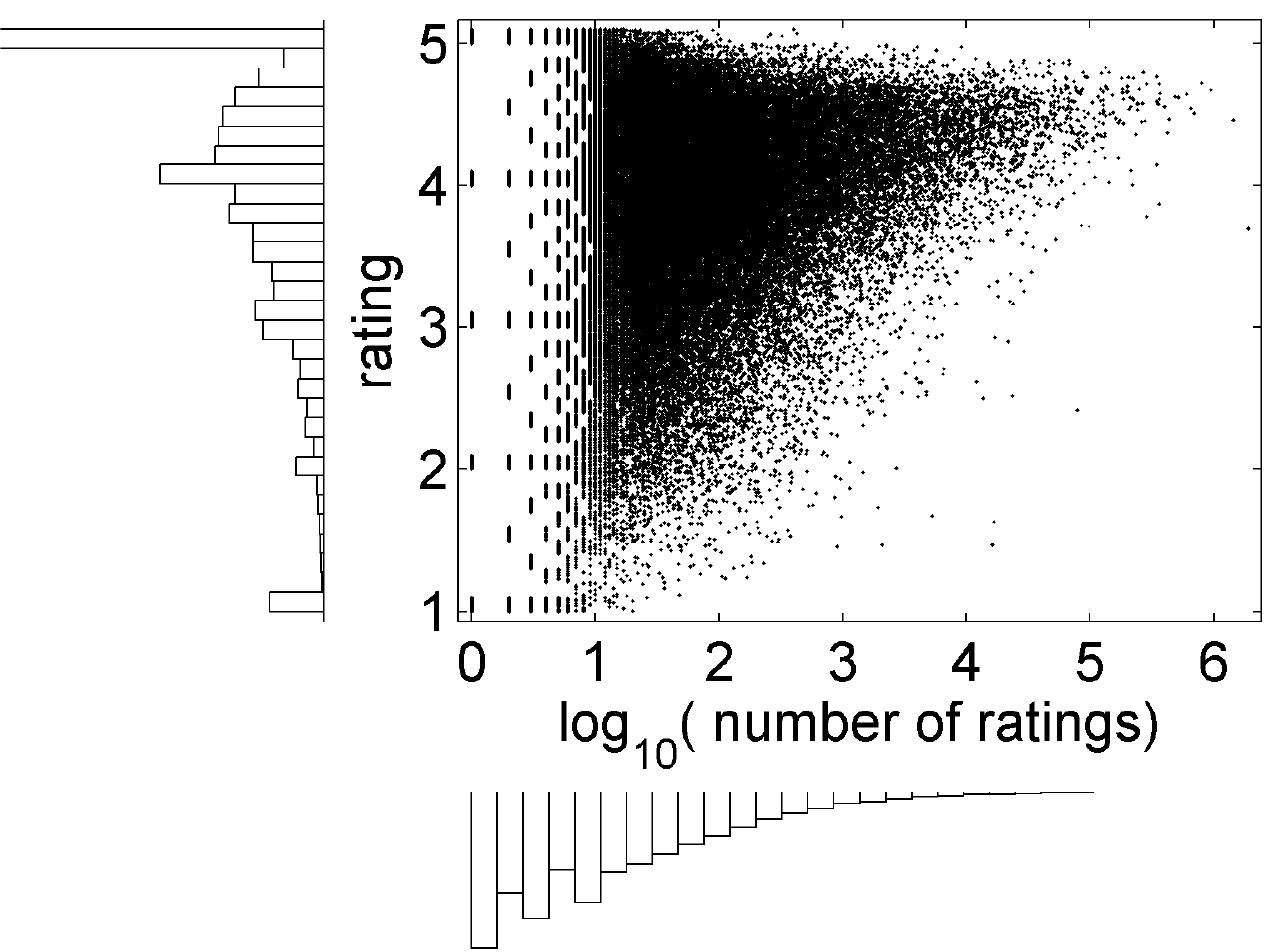}
    } 
\caption{
{\it Left:} Cumulative distribution of Android application prices. The $y$-axis starts at 60\%, and we cut off the distribution at 98.8\% probability mass.
{\it Right: }
Average Android rating scores against the log of the total number of ratings.  Each dot corresponds to a single application. We used a logarithmic scale for the number of user ratings due to its large dynamic range, and we discarded all applications with 0 ratings.
The distribution of average ratings and number of ratings are next to each respective axis.
}
\end{figure*}

\paragraph{Price Distribution.} Figure~\ref{fig_prices} depicts the cumulative
price distribution for Android applications. (Facebook applications are free, so
we do not provide a price analysis for them.) Applications cost up to \$200, and
a majority of all applications are free. Globally, the price
distribution has the most probability mass at low prices and a long tail at high
prices. This is evident due to the almost monotonically decreasing slope of the
cumulative distribution.

We find that developers are strategic in their pricing of applications. This is
evidenced by a strong localized effect that manifests itself at prices just
under each full dollar amount. This effect is the most noticeable at the \$1
area: very few applications cost between \$0.01 and \$0.70, but many
applications cost between \$0.70 and \$1. In the \$2 area, this effect is
apparent again, and this repeats on a dollar interval. The magnitude of this
effect monotonically decreases with increasing price. An exception to this trend
can be seen around \$10, which may be a psychologically significant price due to
its extra digit.

\paragraph{User Ratings.} In Figure~\ref{fig_ratings}, we plot the average user
rating score against the number of user ratings. The logarithm of the number of ratings has a long-tail distribution, which shows that most applications have
very few ratings whereas only a few applications have more than $10^5$ ratings.
Except for two significant peaks at scores of 1 and 5, the distribution of
average ratings follows a Gaussian distribution centered at a score of 4. The peaks are due to the large number of applications that only have a few ratings, all of which are either 1 or 5 stars.

By examining the relationship between average score and the number of ratings,
we can see that applications with higher scores tend to have more ratings. This
suggests that high-quality applications are downloaded and rated more often than
low-quality applications. Additionally, extreme average scores (1 or 5) are only
obtained for applications with a small number of ratings, typically less than
ten. These two observations indicate that the score alone is an insufficient
measure of quality; one also must take the number of ratings into account.

\begin{figure*}[htb]
\centering
 
     \includegraphics[width=0.4\textwidth]{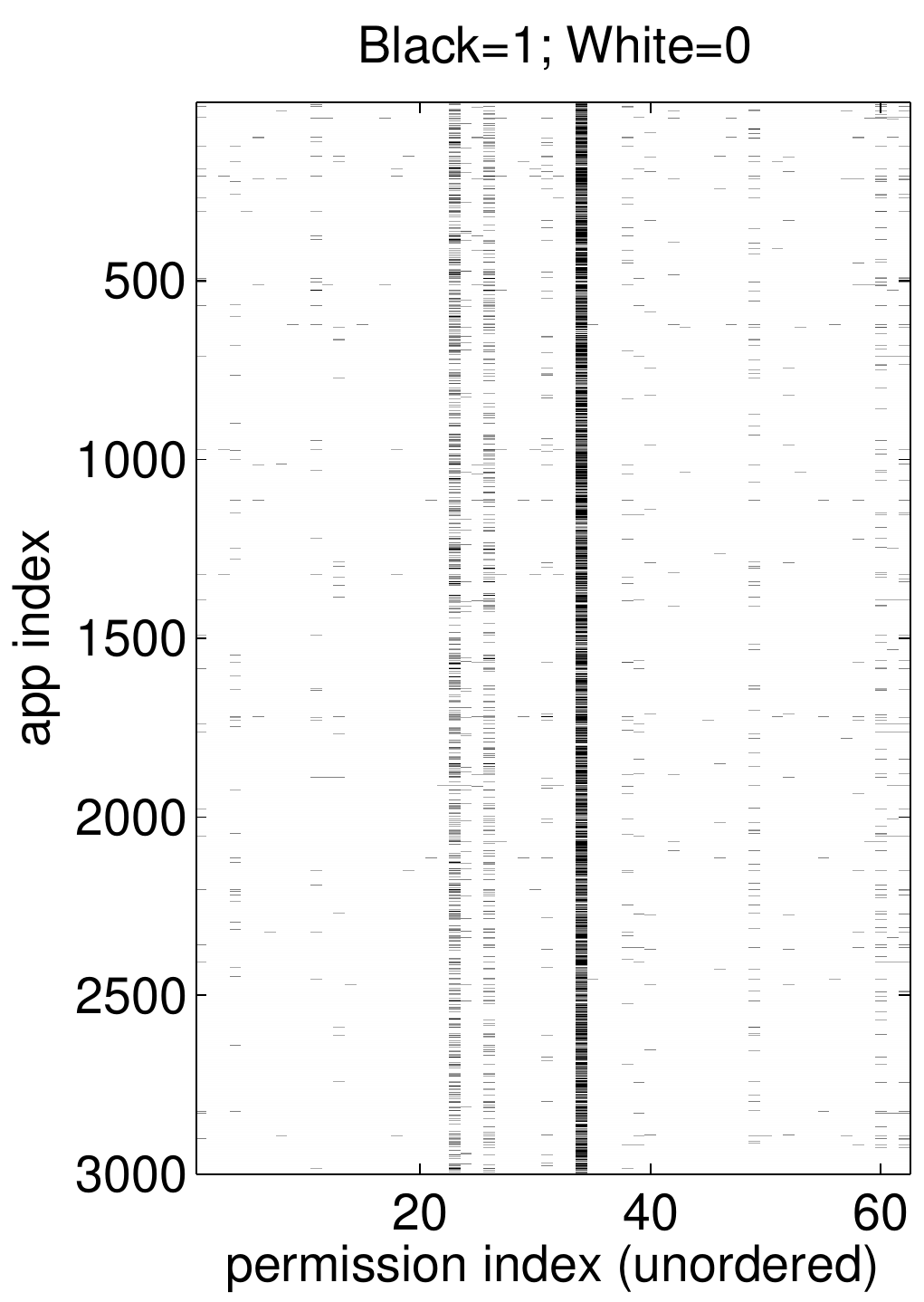}
\caption{ Permission requests from 3000 randomly sampled Facebook applications. Permissions are randomly ordered. A black dot at matrix entry $(i,j)$ means that Application $i$ request Permission $j$.
\label{fig_FB}
}
\end{figure*}

\section{Pattern Mining Technique}
\label{sec_technique}

Our goal is to infer statistically significant permission request patterns from the set of all applications' permission requests. Let $N$ be the number of applications, and let $D$ be the total number of possible permissions. We can then represent the dataset of applications' permission requests as a binary matrix $\mathbf x \in \{0,1\}^{N\times D}$.  The entry $ x_{i d}=1$  means that application $i$ requests permission $d$. The row  $\mathbf x_{i\allInd}\in \{0,1\}^{D}$ represents all permission requests of application $i$.

Given this matrix as an input, we want to find the two following binary matrices: the matrix $\mathbf u$ that encodes which permissions are frequently requested together (the permission request patterns), and the matrix $\mathbf z$ that encodes which applications share the same permission request patterns. Let $K$ be the number of patterns found, then $\mathbf z \in \{0,1\}^{N\times K}$ and $\mathbf u \in \{0,1\}^{K\times D}$.

These two matrices represent an approximate factorization of the input matrix via the Boolean matrix product, where the Boolean product $\mathbf{c} = \mathbf{a} \BMP \mathbf{b}$ of two matrices $\mathbf{a}\in\left\{0,1\right\}^{N\times K}$ and $\mathbf{b}\in\left\{0,1\right\}^{K\times D}$ is defined as (see \cite{MAC_jmlr} for more detailed introduction):
\begin{equation}
\label{eq:BoolMatProd}
c_{id}=\bigvee_{k = 1}^K \left(a_{i k}\wedge b_{k d}\right) \ .
\end{equation}

Having introduced these concepts, we can rephrase our goal as finding a factorization $(\mathbf z^*, \mathbf u^*)$ that
approximates the permission request matrix $\mathbf x$ up to such residuals that exhibit no statistically significant pattern, i.e.,
%\begin{equation}
$
\mathbf x \approx \mathbf z^* \BMP \mathbf u^* \ .
$
%\end{equation}

\subsection{Model and Algorithm} \label{sec_algo}
We employ a probabilistic model for binary matrix factorization \cite{MAC_jmlr}. This model takes a binary matrix $\mathbf x$ and computes the likelihood that a given factorization $(\mathbf z, \mathbf u)$ represents the statistically significant patterns of $\mathbf x$.
We then tweak the entries of $(\mathbf z, \mathbf u)$ to maximize the likelihood of $\mathbf x$. The outcome is a factorization $(\mathbf z^*, \mathbf u^*)$ that does not necessarily provide an exact representation of $\mathbf x$. This model was originally derived as an approach to the role mining problem \cite{Kuhlmann,Vaidya:2010:RMP:1805974.1805983}, where the goal is to identify
roles to configure role-based access control (RBAC)~\cite{rbacOrig}.  Sets of
permission requests can be equated to roles.

The likelihood function in  \cite{MAC_jmlr} explicitly models the data as a probabilistic mixture of signal and noise. In the context of mining permission requests, signal corresponds to significant patterns of permission requests, and noise corresponds to the residuals when fitting these patterns to the data. Each permission request of each application is assumed to follow either the signal distribution $ p_S\left(x_{i\sep d} \left.\right|      \mathbf{z},\boldsymbol{\beta} \right) $ with probability $(1-\epsilon)$, or a random Bernoulli distribution
$p_N \left(x_{i\sep d}\left.\right|r\right)$ with probability $\epsilon$.
The signal distribution is
\begin{eqnarray}
 p_S\left(x_{i\sep d} \left.\right|      \mathbf{z},\boldsymbol{\beta} \right)
  &\!\!=\!\!&
\left[1-\prod_{k=1}^K\beta_{kd}^{z_{ik}}
 \right]^{x_{id}} \left[\prod_{k=1}^K\beta_{kd}^{z_{ik}}
\right]^{1-x_{id}} \label{eq_signal}
\end{eqnarray}
Here, the Boolean permission requests $u_{kd}$ assigning permission $d$ to pattern $k$  are modeled by the probability that they are $0$, i.e. $\beta_{kd}:=p(u_{kd}=0)$. Assuming that the individual entries $x_{i\sep d}$ are independent, (\ref{eq_signal}) is  a modified Bernoulli distribution $B(x_{i\sep d}\vert q_{id})$ with Bernoulli parameter $q_{id}=\prod_{k=1}^K\beta_{kd}^{z_{ik}} $.

The noise distribution is a Bernoulli distribution 
$
 p_N \left(x_{i\sep d}\left.\right|r\right)
  =
   r^{x_{i\sep d}}\left(1-r\right)^{1-x_{i\sep d}}
$.
This means that if an application's permission request $x_{i\sep d}$ is generated from the noise distribution, then it is sampled from a biased coin flip and with probability $r$ application $i$ requests permission $d$.

Finally, the complete likelihood function %$p\left(\mathbf{x}\left.\right|\mathbf{z},\boldsymbol{\beta}, r,    \epsilon \right)$
 %together with
 is a mixture of the noise probability distribution and the signal probability distribution:
\begin{eqnarray}
 p\left(\!\mathbf{x}\vert\mathbf{z},\boldsymbol{\beta}, r,
    \epsilon \!\right)
 &\!\!\!\!=\!\!\!\!&
 \prod_{i=1}^N \prod_{d=1}^D \! \left(
    \epsilon p_N\! \left(x_{i\sep d}\vert r\right)
    + \!(1\!-\!\epsilon)  p_S\!\left(x_{i\sep d}\vert     \mathbf{z},\boldsymbol{\beta} \right) \right)
\nonumber
\end{eqnarray}

The parameters $(\mathbf{z},\boldsymbol{\beta}, r, \epsilon )$ of this distribution can be optimized by an annealed expectation-maximization algorithm~\cite{MAC_jmlr}. This algorithm alternates between updating the individual parameters and, after each iteration, reducing the computational temperature, ultimately forcing the probabilistic parameters $\beta_{kd}\in[0,1]$ towards $0$ or $1$. The algorithm terminates when the parameter updates become negligible or when a predefined temperature is reached.

The optimization procedure outputs
the set of parameters $(\mathbf{z}^*,\mathbf u^*, r^*,    \epsilon^* )$.
The Boolean matrix product of $\mathbf{z}^*$ and $\mathbf u^*$ provides an approximation of the input matrix, i.e. $\mathbf x \approx \mathbf z^* \BMP \mathbf u^*$.
The two Bernoulli parameters $r^*$  and  $\epsilon^*$ model the residuals. Here $\epsilon^*$ is the probability that a permission request is not determined by $\mathbf z^* \BMP \mathbf u^*$, and $r^*$ is the probability that such a permission request is 1. (A Boolean 1 means that the permission is requested by the particular application.) %With probability $(1-r^*)$ a noisy permission request is 0.

The factorization $(\mathbf z^*, \mathbf u^*)$ can be interpreted in several ways. The matrix $\mathbf u^*$ provides the assignment of permissions to permission request patterns, where one permission can appear in multiple patterns.
The matrix $\mathbf z^*$ assigns these patterns to applications. The disjunction of all permission request patterns of an application constitutes the set of permissions that an application requests.
At the same time, $\mathbf z^*$ is a clustering of applications into groups of applications that request the same set of permission patterns.

\subsection{Selecting the Number of Patterns}\label{sec_instab}

The number of patterns $K$ must be provided as an input to the algorithm described in Section~\ref{sec_algo}. This is an important parameter. For a simple dataset, a small number of patterns might suffice to capture all relevant information, whereas an overly high number of patterns could lead to overfitting. On the contrary, a complicated dataset might require a large number of different patterns; too few patterns could lead to missing significant individual permissions. % In essence, we face the model-order selection problem that usually arises in unsupervised learning problems.

There are several related heuristics for  this model-order selection problem \cite{mtc_ECML,Lange2004Stability} and we approach it by carrying out an instability
analysis~\cite{Lange2004Stability}. The instability measure is based on the
requirement that a clustering solution is reproducible. This means that a model
should ideally find the same patterns or clusters in different random subsets of
the data. 
To quantify the degree to which this
requirement is satisfied, we subdivide $\mathbf x$ into
two independent and identically distributed data sets $\mathbf x^{(1)}$ and
$\mathbf x^{(2)}$.
Then we run the factorization algorithm on both datasets, resulting in two clusterings of the applications: ${\mathbf z}^{(1)}$ and ${\mathbf z}^{(2)}$.
A good clustering algorithm will produce two similar clusterings. If the clusterings are exactly the same, then the algorithm is able to precisely find the patterns in the data that are significant (i.e., repeatable over the two subsets).  If the clusterings differ a lot, then the algorithm has learned patterns that are random noise (i.e., they do not repeat over random subsets of the data).

It is difficult to compare clustering results for two different datasets. The solution taken in \cite{Lange2004Stability} is to train a classifier $\phi^{(1)}$ on the first data
set $\mathbf x^{(1)}$ and the cluster assignments $\mathbf z^{(1)}$ and then to predict the cluster assignment of each object $i$ from the second dataset $\tilde{\mathbf z}^{(2)}_{i\sep \allInd}:= \phi^{(1)}\! \left( x^{(2)}_{i\sep \allInd}\right)$\vspace{-0.4em}. Thereby, the binary vector $\tilde{\mathbf z}^{(2)}_{i\sep \allInd}$ indicates the assignment to all clusters (the placeholder index $\allInd$ runs from $1$ to $K$) and the object can be assigned to multiple clusters at once (the sum $\sum_k \tilde{\mathbf z}^{(2)}_{i\sep k}$ can be larger than 1).

The numbering of clusterings is arbitrary and could incidentally differ between the solutions. Therefore, one must find the
permutation $\pi$ that minimizes the deviation. Taking this into account, the instability is
defined as
\begin{equation}
s := \frac{\left|\lsetset\right|}
      {\left|\lsetset\right|-1}
    N^{-1}
    \min_{\pi \in P_K}\left\{
      \sum_{i=1}^N
      \eeins
      \left\{
      \tilde{\mathbf z}^{(2)}_{i\sep \allInd}
        \neq \mathbf z^{(2)}_{i\cdot}
        \right\}
    \right\} \ .
\end{equation}
The predicate $ \eeins
      \left\{
      \tilde{\mathbf z}^{(2)}_{i\sep \allInd}
        \neq \mathbf z^{(2)}_{i\cdot}
        \right\}$ holds if the cluster assignment vectors of object $i$ $\tilde{\mathbf z}^{(2)}_{i\sep \allInd}$ and $\mathbf z^{(2)}_{i\cdot}$ differ with respect to at least one cluster.
      The factor
$\frac{\left|\lsetset\right|} {\left|\lsetset\right|-1}$ is needed
to compare clustering solutions with different numbers of clusters %(or patterns in our case)
 in a fair way. This can be seen by considering completely random clusterings which should ideally lead to the same instability for different numbers of clusters.
Let $\lsetset$ be the number of possible assignment vectors $\mathbf z_{i\cdot}$ ($2^K$ in the case of $K$ patterns), then randomly assigning data items with the $|\lsetset|$ assignment vectors
yields a stability of $\frac1{|\lsetset|}$.

\begin{figure}[htb]
    \begin{minipage}[c]{0.59\linewidth}
\centering
\includegraphics[width=1\textwidth]{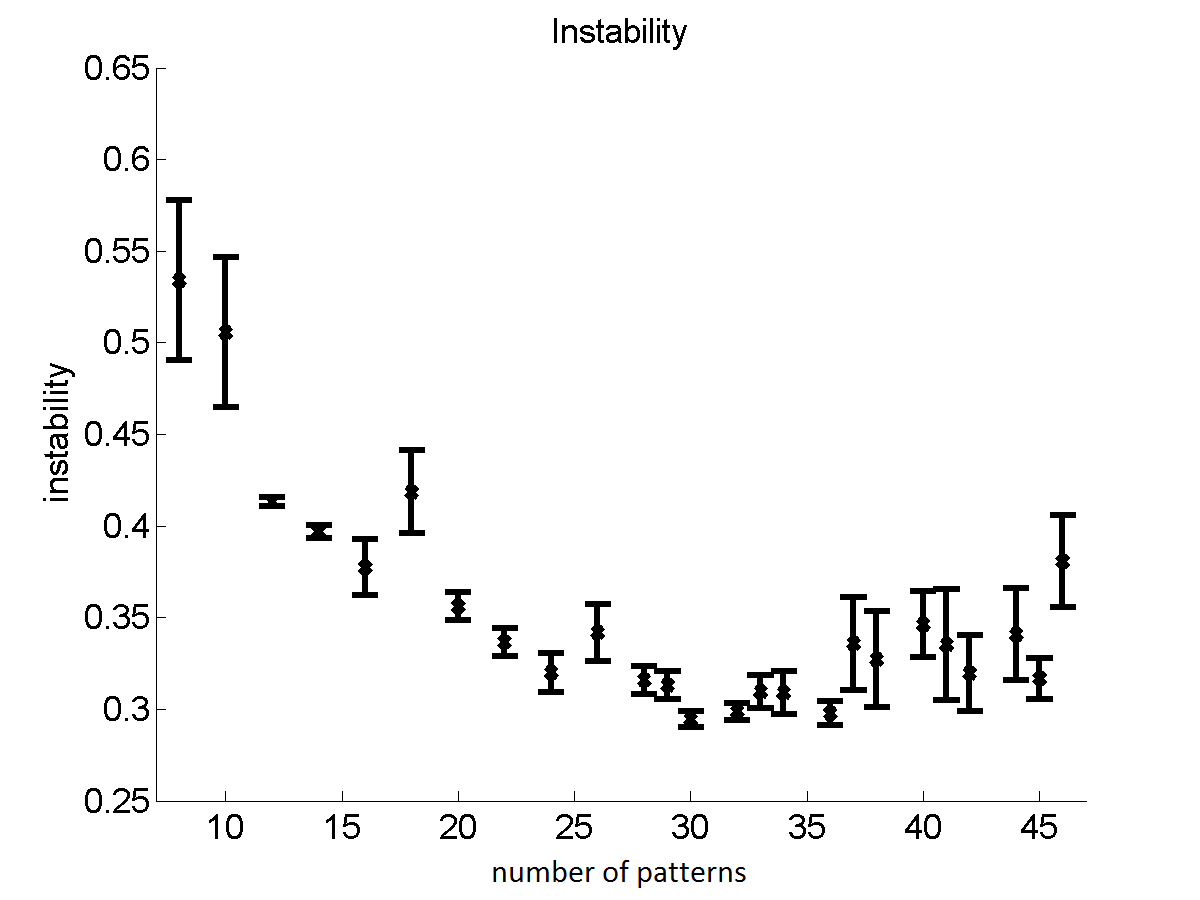}
    \end{minipage}
    \hfill
    \begin{minipage}[c]{0.4\linewidth}
     \caption{Trend of the clustering instability as a function of the number of permission request patterns for Android.
\label{fig_instab} }
    \end{minipage}
\end{figure}

As an example, Figure~\ref{fig_instab} depicts the outcome of the instability analysis of Android permission requests. We computed the instability for different numbers of permission request patterns between $K=8$ and $K=46$. For each number of patterns $K$, we computed the instability based on five random partitions of the dataset into two subsets $\mathbf x^{(1)}$ and $\mathbf x^{(2)}$. We then plotted the median instability along with error bars that account for the standard deviation. As one can see, the instability is minimized at 30 permission request patterns.
In the underfitting phase ($K<30$), the instability is higher because the significant patterns that have been left out are randomly assigned to some of the existing patterns. This division of patterns is arbitrary, and thus not reproducible over the two random datasets. In the overfitting phase ($K>30$), the extra patterns are fitted to random data-structures resulting from noise that do not reproduce over $\mathbf x^{(1)}$ and $\mathbf x^{(2)}$, thus leading to a higher instability. In this phase, the variance of the instability also increases, since misinterpreting the random structures strongly depends on the particular noise realizations of the two subsets of the data.
Following the results of this analysis, we selected $K=30$ patterns to fit the Android dataset and $K=5$ patterns to fit the Facebook dataset.

\section{Experiments}\label{sec_Exps}

We mine 
 patterns of permission requests from
high-reputation Android and Facebook applications.
Section~\ref{sec_PermGroups} details our methodology and provides an overview
of the permission request patterns that we found. We then consider how
permission request patterns differ between high- and low-reputation
applications (Section~\ref{sec_HiReputationGroups}) and find that the patterns
can be used as part of a risk metric for new, unknown applications. We also
consider whether request patterns are related to Android categories
(Section~\ref{sec_categories}).

\subsection{Overview of the Permission Request Patterns}
\label{sec_PermGroups}

We apply the techniques described in Section~\ref{sec_technique} to the permission requests of high-reputation Android and Facebook applications.

\subsubsection{Methodology} We train our model on {\it high-reputation applications}, defined as applications with average ratings of 4 or higher and at least 100 user ratings. As discussed in Section~\ref{globalStats}, our reputation metric must combine average ratings and the number of reviews because the average rating by itself is an unreliable measure. This quality criterion yields 11,554 Android applications and 1,998 Facebook applications. For Android, we reserved $2,000$ applications to use as a test set and trained on the remaining 9,554 high-reputation applications. For Facebook, we reserved $400$ test applications and trained on 1,598 applications.

In order to perform the instability analysis (Section~\ref{sec_instab}), we
randomly subdivided the training sets into halves. The outcome of this
analysis led us to select $K=30$ patterns for the Android dataset and $K=5$
for 
Facebook. 
 We then used the entire training set to fit the
probabilistic model.

\begin{figure}[tb]
\centering
\subfloat[PRPs]  {
\centering
\includegraphics[width=0.43 \textwidth]{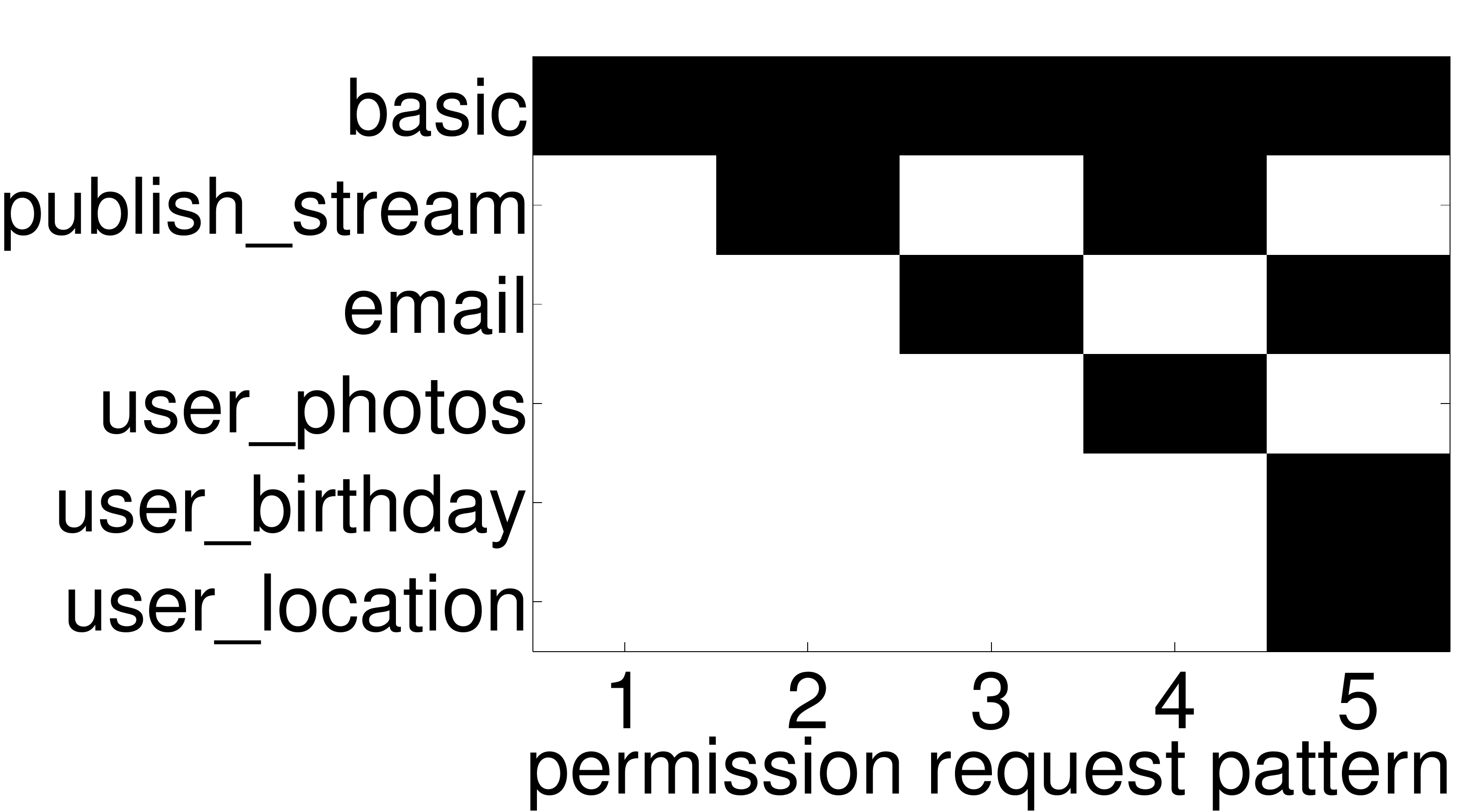}
\label{fig_FBPermGroups}
}
\ \ \ \ \ \ \ \
 \subfloat[Pairwise Conditional Probablities]  {
\centering
\includegraphics[width=0.43\textwidth]{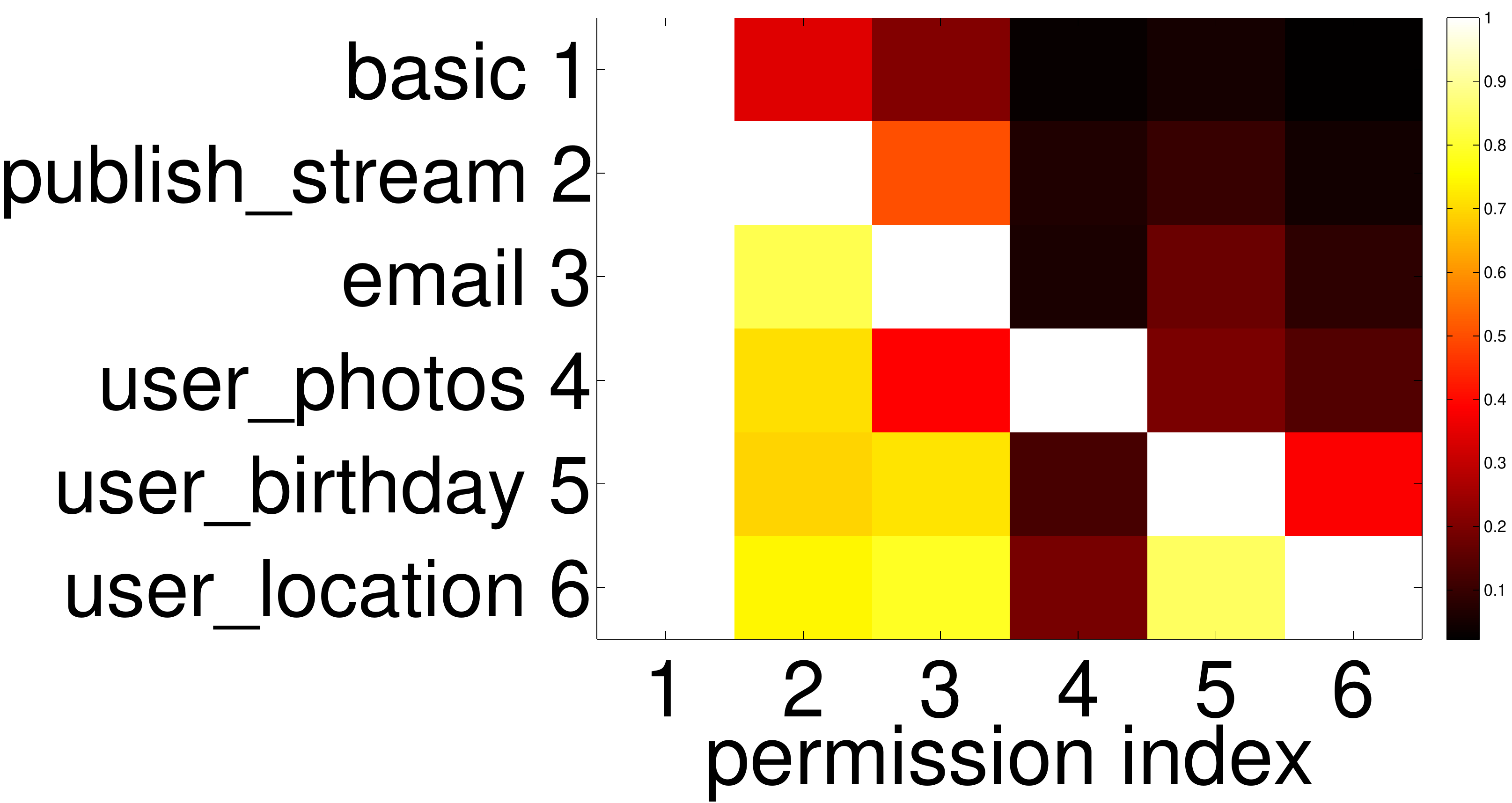}
\label{fig_FBasso}
}
\caption{PRPs and pairwise conditional probabilities for Facebook. Each column in Fig.~\ref{fig_FBPermGroups} is a PRP;  a black entry in the matrix means that the permission is part of the PRP; the patterns are sorted from left to right with decreasing popularity. Fig.~\ref{fig_FBasso}: bright colors indicate a high conditional probability.
\label{fig_permgroupsAndAssoFB}
}
\end{figure}

\begin{figure}[htb]
    %\begin{minipage}[c]{0.65\linewidth}
\centering
\includegraphics[width=0.8\textwidth]{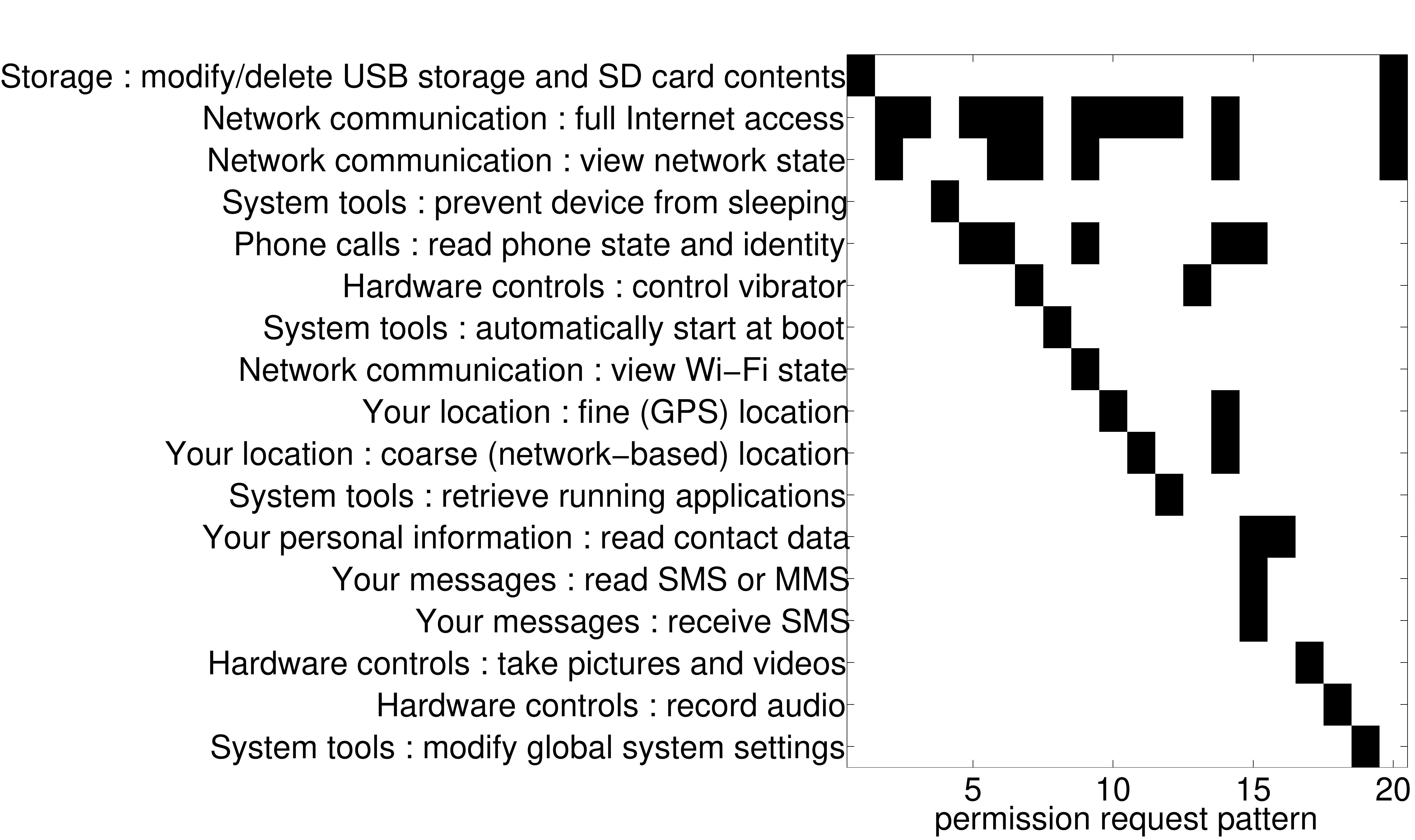}
 %   \end{minipage}
    %\hfill
    %\begin{minipage}[c]{0.34\linewidth}
     \caption{The permissions included in the 20 most popular Android permission request patterns (PRPs).  The fraction of applications requesting the PRP decreases from left to right.
\label{fig_AndoidPermGroups} %\vspace{-1.5em}
}
    %\end{minipage}
\end{figure}

\begin{figure}[htb]
  %  \begin{minipage}[c]{0.65\linewidth}
\centering
\includegraphics[width=0.8\textwidth]{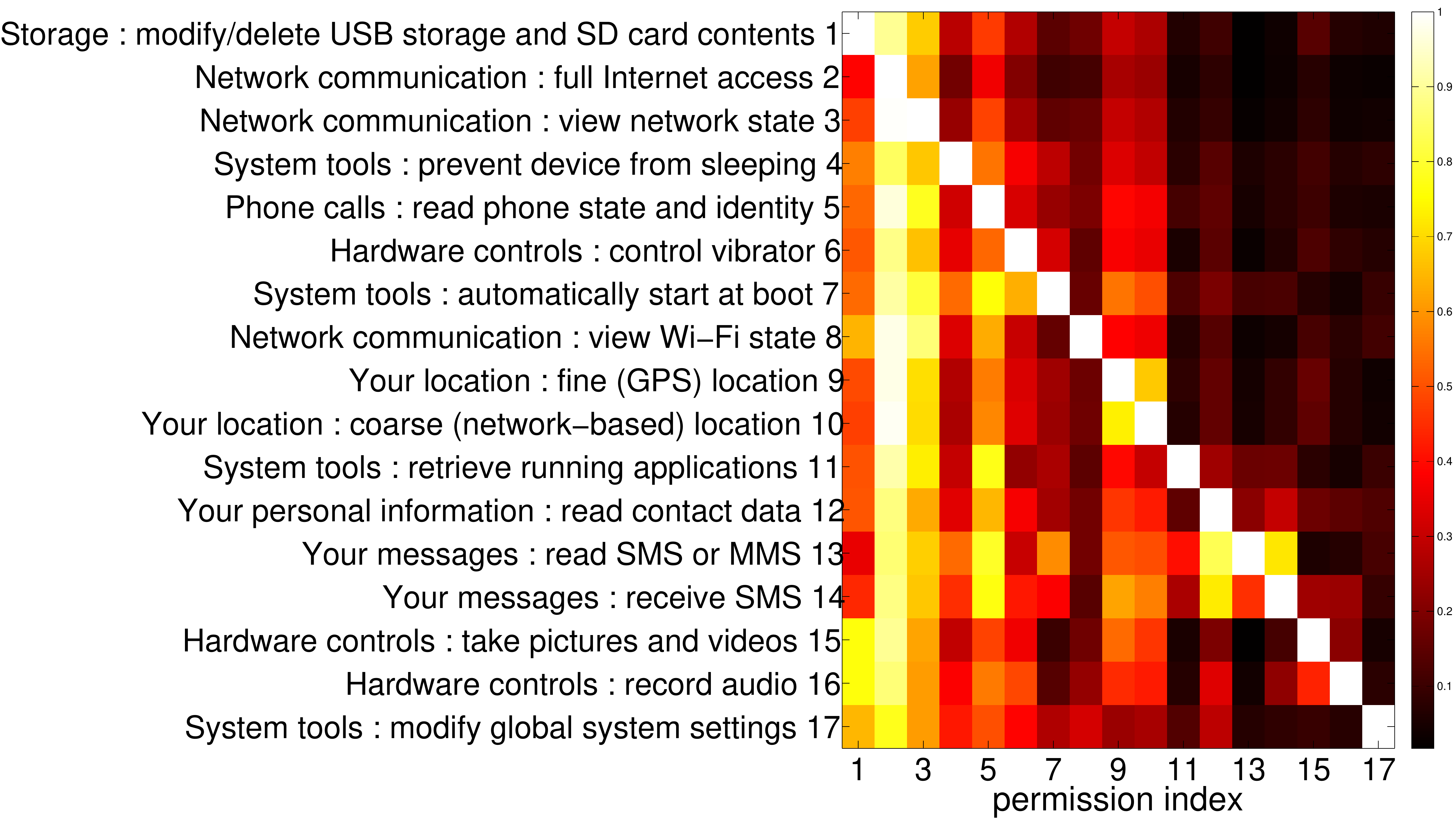}
 %   \end{minipage}
  %  \hfill
 %   \begin{minipage}[c]{0.34\linewidth}
     \caption{Pairwise conditional probabilities for the Android permissions from the 20 most popular patterns.
\label{fig_Andrasso} 
%\vspace{-1.5em}
}
 %   \end{minipage}
%\vspace{-1.5em}
\end{figure}

\subsubsection{Results} The Facebook dataset has a simple and clean
structure. We found five permission request patterns (PRPs) for Facebook, which
suffice to cover the requested permissions of most Facebook applications. Figure~\ref{fig_FBPermGroups} depicts the five PRPs.  
This simple structure might be due to the fact that the permissions  requested after installation are not accounted for in the data, as explained in Section~\ref{sec_datacollect}.

The
Android dataset is more complex: our mining technique identified 30 PRPs for
Android. The most common PRPs for Android are shown in
Figure~\ref{fig_AndoidPermGroups} and
Table~\ref{tab_topPerms}. Following the notation used
in Section~\ref{sec_algo}, Figures~\ref{fig_FBPermGroups}  and \ref{fig_AndoidPermGroups} are the transposed
matrices $\mathbf u^*$ that assign permission patterns to permissions. We
sorted the patterns by frequency such that PRP1 is the most common pattern,
PRP2 is the second most-requested pattern, etc.

PRPs are not disjoint: permissions can be members of
multiple patterns, and applications can request multiple patterns.
Consequently, most patterns only include a small number of permissions. A PRP
with only one permission reflects the fact that the permission is requested
very frequently, but not always together with the same other permissions.

In order to explain why particular permissions appear in the same pattern, we can consider the pairwise conditional probabilities. For two permissions $s$ and $t$, the conditional probability is empirically estimated as
\begin{equation}
 p_{st} :=p(x_{s}=1 \vert x_{t}=1) = \left( \sum_{i=1}^N x_{i t} \right)^{-1} \sum_{i=1}^N x_{i s} x_{i t} 
 \label{eq_PCP}
\end{equation}
 If $p_{st}=1$, then $s$ is requested whenever $t$ is requested.
This score is not symmetric, i.e., $p_{st} = p_{ ts} $ is not necessarily true.
We plot these scores for Android and Facebook in the heatmaps in Figure~\ref{fig_FBasso} and Figure~\ref{fig_Andrasso}, respectively. The brighter the color, the higher the conditional probability. It is apparent that pairs of permissions with a high conditional probability often end up in the same pattern. We found that clusters of permissions with a high pairwise conditional probabilities can not emerge by chance. We detail this analysis in Section~\ref{sec_discussion}.

\begin{figure*}[htb]
\centering
\includegraphics[width=0.474\textwidth]{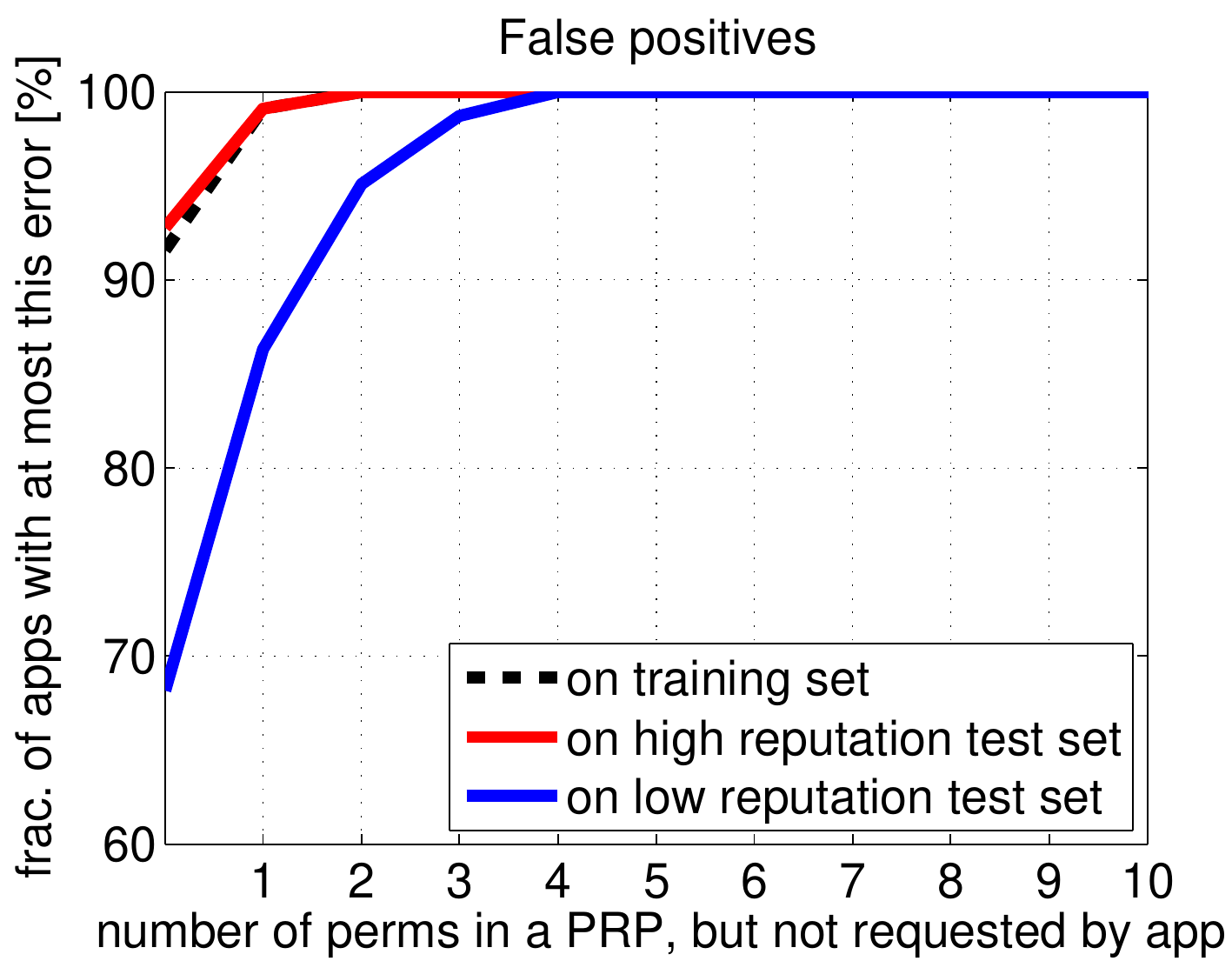}
\ \ \ \ \ \ \ \
\includegraphics[width=0.45\textwidth]{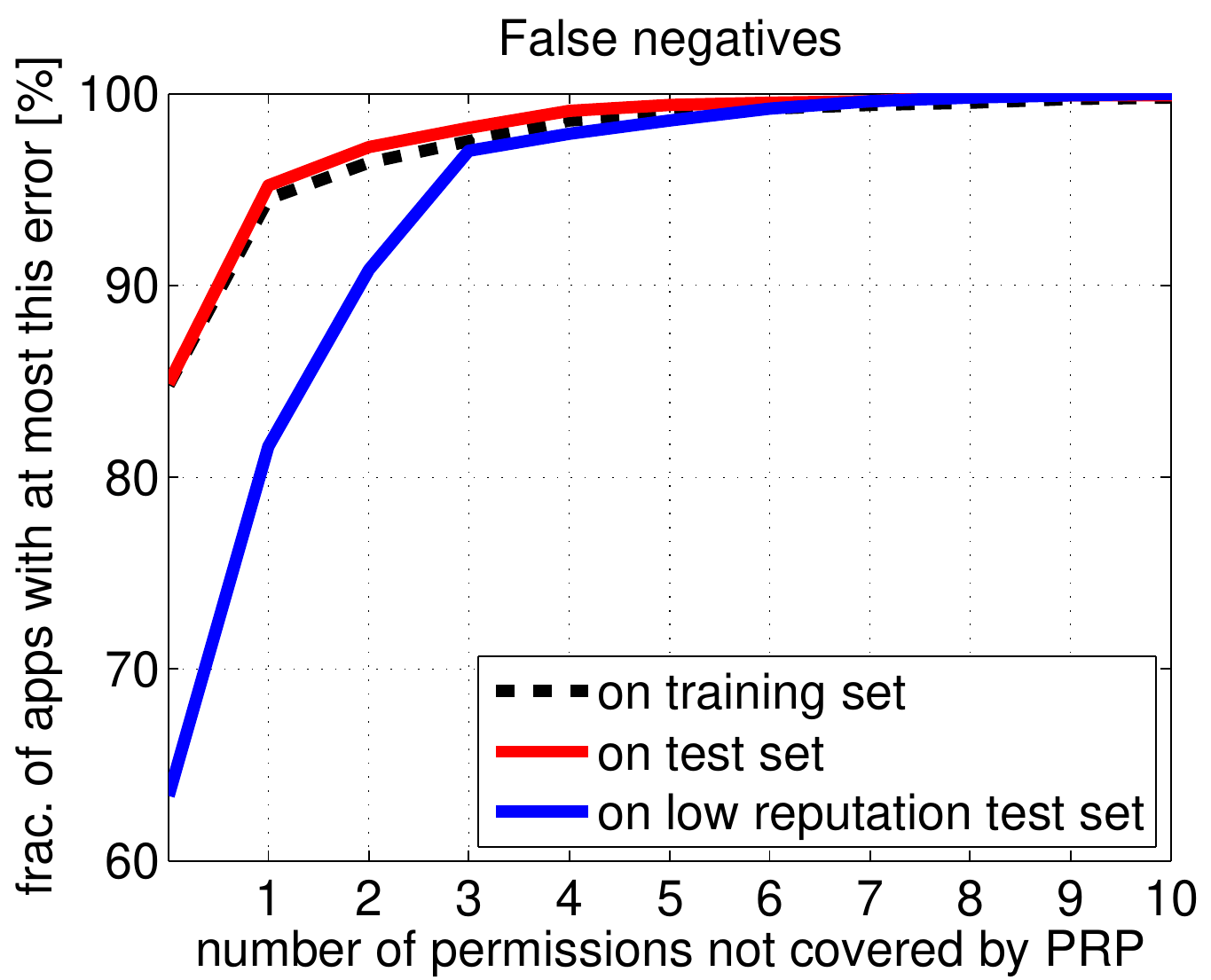}
\caption{Fraction of Facebook applications with a particular error rate.
\label{fig_fnfpFB}
}
\end{figure*}

\subsubsection{Evaluation}

\begin{figure*}[htb]
\centering
\includegraphics[width=0.474\textwidth]{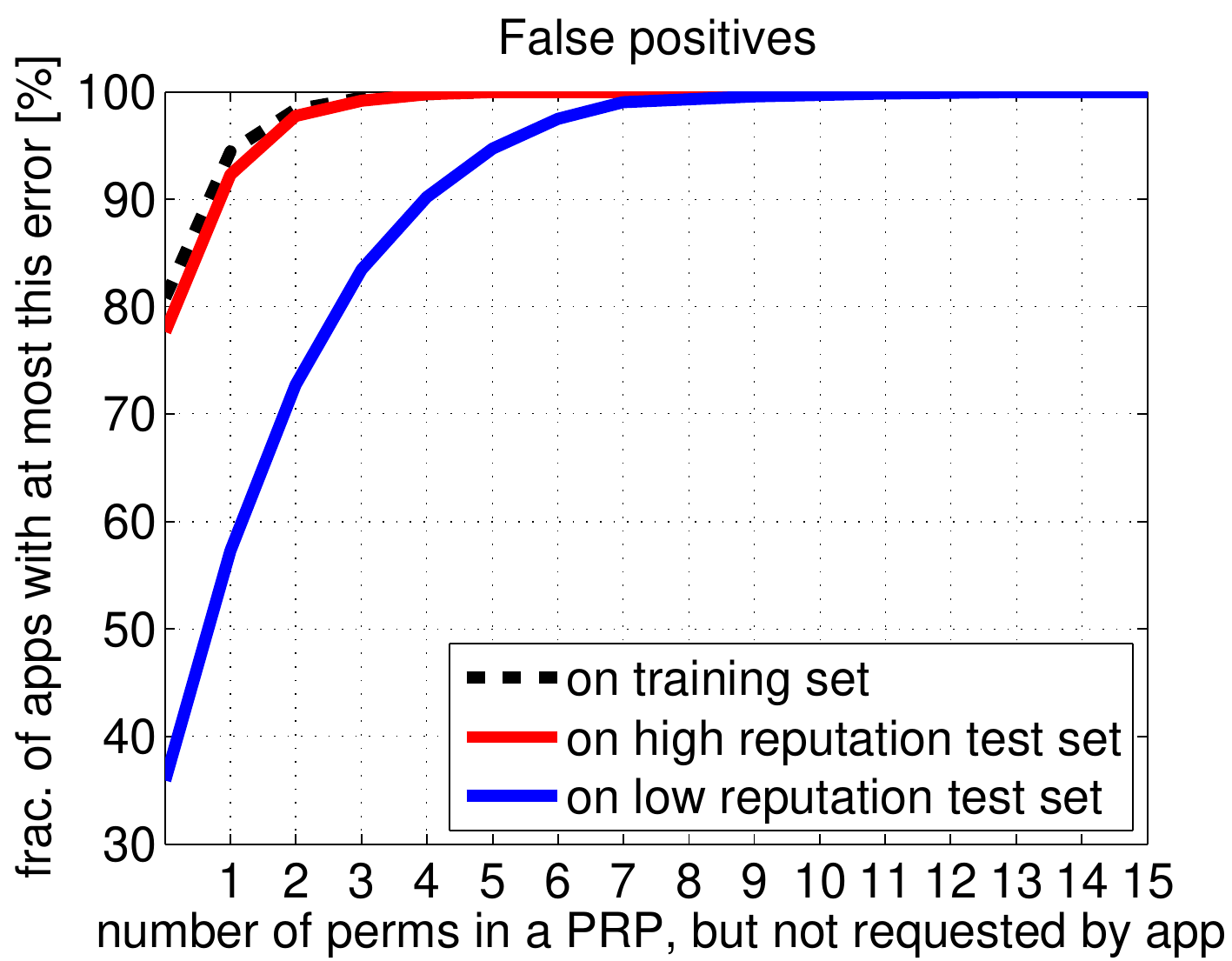}
\ \ \ \ \ \ \ \
\includegraphics[width=0.46\textwidth]{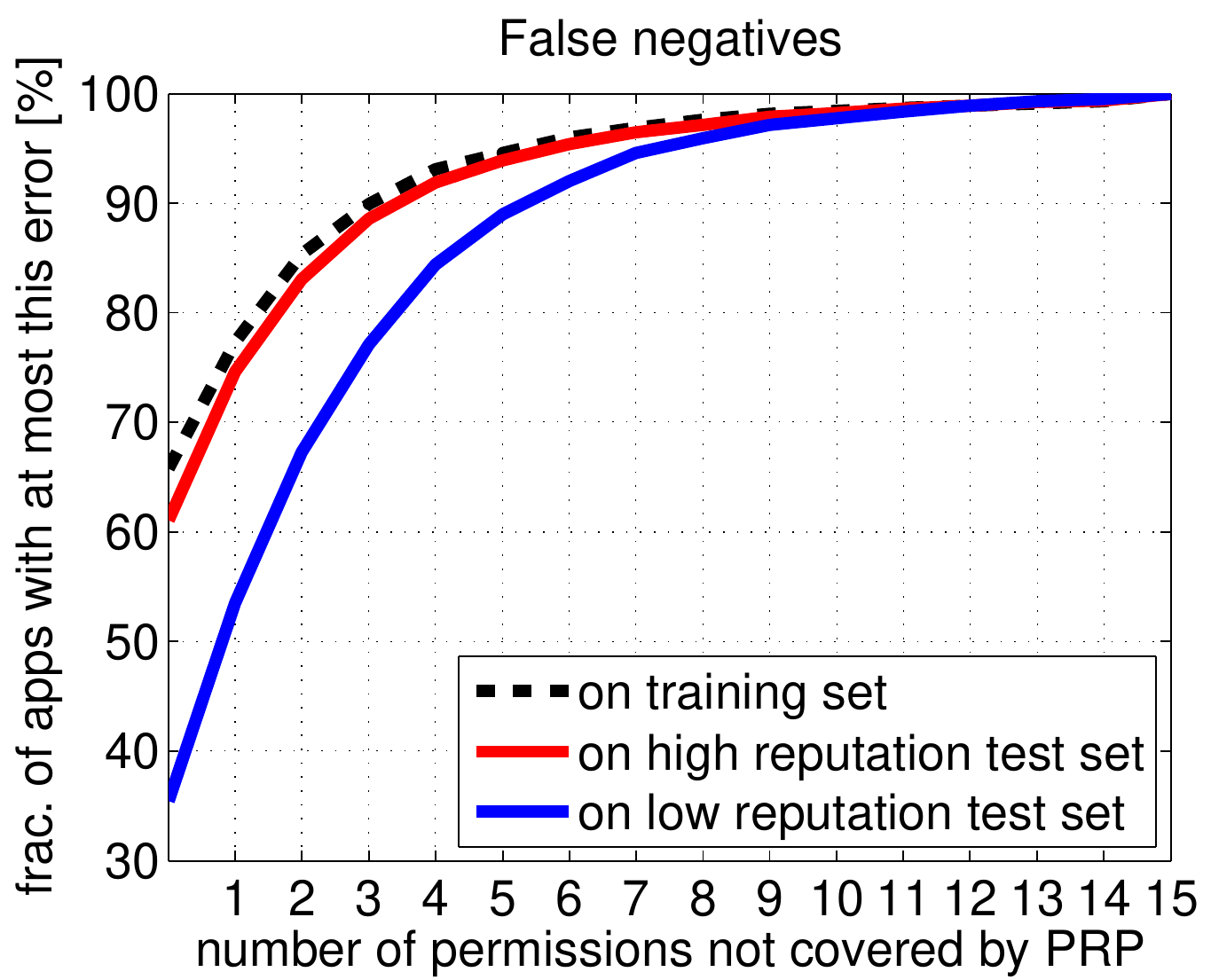}
\caption{Fraction of Android applications with a particular error rate.
\label{fig_fnfpANDR}
}
\end{figure*}

We evaluate how well the permission request patterns generalize and fit the data by considering the false positive and false negative rates. Ideally, the patterns should yield low false positive and false negative rates for both the training sets and the test sets. A {\it false positive} occurs when an application is assigned to a PRP without having all of the permissions associated with the PRP, and a {\it false negative} occurs when an application's permission requests are not covered by any of the PRPs that the application is assigned to. Consequently, we define the false positive rate $f_p$ as the average number of permissions that are incorrectly assigned to applications, and the false negative rate $f_n$ as the average number of permissions per application that are not covered by PRPs. Given the reconstruction matrix $\mathbf x^* =\mathbf z^* \BMP \mathbf u^*$ and an indicator $\eeins\{t\}$ that is 1 if the predicate $t$ is true and 0 otherwise:
\begin{eqnarray}
f_n \!\! &:=&\!\! N^{\!-\!1}\sum_{i=1}^N \sum_{d=1}^D \eeins \{ (x_{id}\! - \! x^*_{id})\! =\! 1\} \ ,
\\ 
f_p\!\!& :=&\!\! N^{\!-\!1}\sum_{i=1}^N \sum_{d=1}^D \eeins \{ (x_{id}\! - \! x^*_{id}) \!=\! -1\} \ .
\end{eqnarray}
We depict the cumulative false positive and false negative rates in Figures~\ref{fig_fnfpFB} and~\ref{fig_fnfpANDR}. (Please note that we cut off the y-axis at different values to best resolve the dynamic range for each experiment.)  Our evaluation focuses on the error rates for the training set and high-reputation test set in Figures~\ref{fig_fnfpFB} and~\ref{fig_fnfpANDR}; we discuss the error rates for the low-reputation test set in  Section~\ref{sec_HiReputationGroups}.

Among the Facebook applications, 2\% of 
 high-reputation applications have at
least one false positive ($f_p\!>\!0$), and just under 20\% of all high-reputation
applications have false negatives ($f_n\!>\!0$). Thereby, the residuals are
very small: there are almost no applications with more than one false positive
($f_p\!>\!1$) and only 7\% with more than one false negative ($f_n\!>\!1$). The error
rates are higher for Android applications, which reflects the greater complexity
of the Android dataset. Approximately $20\%$ of high-reputation applications
have at least one false positive, and $35\%$ have false negatives.  For both
platforms, the false negative rate is significantly higher than the false
positive rate. The false negative rate can be partially attributed to the very
large number of infrequently-requested permissions; these
unpopular permissions cannot be placed in PRPs, so the applications
that request them cannot be fully fitted with PRPs.

For both platforms, the error rate for the high-reputation test set closely matches the error rate for the high-reputation training set.  This demonstrates that we have likely discovered the true statistical structure of the permission requests for high-reputation applications.

\subsection{Reputation and Risk}
\label{sec_HiReputationGroups}

Our goal is to determine whether permission request patterns can help identify
low-quality applications that users should be cautious about installing. We
trained on high-reputation applications with the assumption that
high-reputation applications are trustworthy, high-quality applications. If this is true, PRPs 
 could be incorporated into a risk
metric for new applications: new applications could be considered riskier or
lower-quality if their permission requests cannot be fitted to high-reputation
PRPs.

We can evaluate the suitability of PRPs for a risk metric by comparing the
false positive and false negative rates for the high-reputation training set,
the high-reputation test set, and a low-reputation test set that includes
applications with fewer than 10 user ratings (regardless of the score).
Figures~\ref{fig_fnfpFB} and~\ref{fig_fnfpANDR} display all three datasets. We
find that low-reputation applications significantly differ from
high-reputation applications in how they request permissions, as evidenced by
the false positive and negative rates.
While the reputable test applications have an error rate that is close to the training error, the unpopular applications have significantly higher residuals, both false positives and false negatives.
This does not indicate that the
low-reputation applications are fraudulent, but it does suggest that
permission requests can be used to help classify an application as high- or
low-reputation. Consequently, we recommend that permission request patterns be
used as part of a risk metric for newly-uploaded applications. For example,
this could be incorporated into a search result ranking algorithm.

\subsection{Android Categories}
\label{sec_categories}

In the Android Market, applications are grouped into 27 application categories
and 8 game categories based on their primary purpose, such as ``Books \&
Reference,'' ``Photography,'' and ``Weather.'' Since categories are roughly
indicative of functionality, we consider whether there exists a relationship
between categories and permission request patterns. If such a relationship
were to exist, then certain patterns would be strongly associated with certain
categories.

\begin{figure}[htb]
\centering
\includegraphics[width=0.7\textwidth]{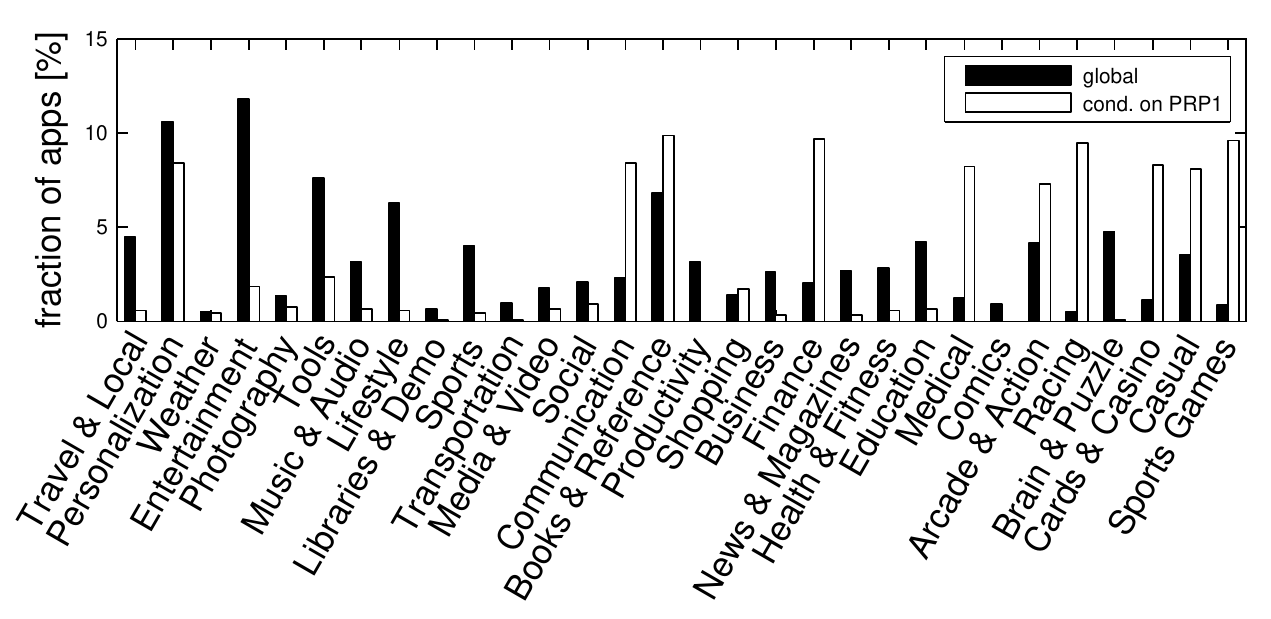}
\caption{Category distribution of all applications and of those that fit PRP1.
If there were no relationship between category and whether an application fits PRP1, both distributions would be the same. However, PRPs do not determine a single category.
\label{fig_tags}
}
\end{figure}

For example, Figure~\ref{fig_tags} illustrates the relationship between PRP1
and categories. Applications that fit PRP1 are more likely to be in certain
categories than general applications are: disproportionately few PRP1
applications are classified as ``Entertainment'' or ``Tools,'' whereas
disproportionately many PRP1 applications are in the ``Communication''
category. Most of the applications that fit PRP1 fall in ten categories. This
indicates that certain categories are more likely than others to contain
applications that request this set of permissions.

To quantify the difference between the distribution of categories across all
applications ($p_g$) and the distribution of categories across applications
that fit a certain PRP ($p_k$), we compute the Kullback-Leibler divergence
$KL(p_g,p_k)$. The depicted example for PRP 1 has $KL(p_g,p_1)=1.52\ \text{bits}$. The KL-divergences of all other PRPs are reported in
Table~\ref{tab_topPerms}. Particular patterns are always assigned to
applications of many categories; i.e., we do not find a one-to-one
relationship between any categories and patterns. However, we do find that the
patterns are informative for the categories of the applications.

\subsection{Sanity Check: Simulated Applications}

In this section, we carry out a sanity check based on simulated permission request data. We want to find out if a set of hypothetical applications that randomly request permissions can exhibit many pairs of permissions that by chance are often co-requested and thus can be misinterpreted as patterns. If this is true, it is invalid to believe that our found patterns reflect true functionality of the applications, as the patterns could as well just occur by chance.

As discussed in Section~\ref{sec_PermGroups} the permission request patterns in the data are due to pairs of permissions that have a significantly higher conditional probability of being requested than other pairs.
This pairwise conditional  probability (PCP), defined in Eq.~(\ref{eq_PCP}), quantifies how likely it is that an application requesting one permission also requests the other.

In order to test the hypothesis that random permission requests can lead to random structures, we carry out the following experiment. For each permission $d$ of the Android dataset, we compute the empirical probability that the permission is requested: $p_d = N^{-1} \sum_{i=1}^N x_{id}$ with $N= 188,389$. We simulate the permission requests of $N$ hypothetical applications. For each application and each permission $d$, we request $d$ with probability  $p_d$. 
Then, we compute the PCP of all pairs of permissions from Eq.~(\ref{eq_PCP}).

\begin{figure}[htb]
\centering
 % \begin{center}
   % \begin{minipage}[c]{0.55\linewidth}
\includegraphics[width=0.48\textwidth]{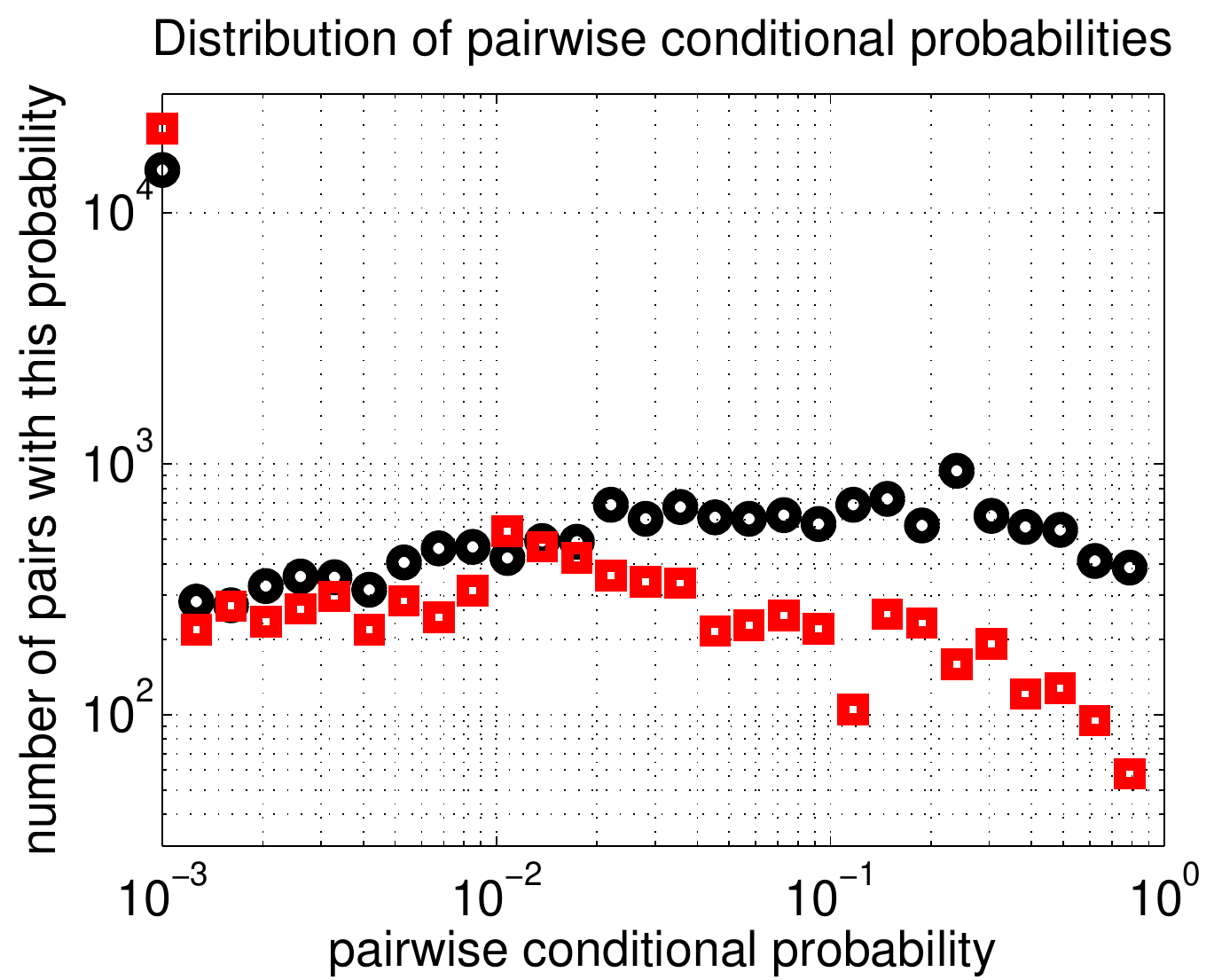}
  %  \end{minipage}
  %  \hfill
   % \begin{minipage}[c]{0.44\linewidth}
     \caption{\label{fig_assoMatsim}Conditional pairwise probabilities (PCP)  of permission requests. Black circles depict the frequencies of PCPs of the Android dataset. Red circles are PCPs for data that has been  simulated according to the permission request probabilities of Android applications. The lowest bin cumulates all permission pairs with lower PCP. }
 %   \end{minipage}
 % \end{center}
\end{figure}

In Figure~\ref{fig_assoMatsim}, we depict the number of permission pairs that have a particular PCP for the real Android dataset and the simulated dataset, respectively. Please note that both axes are logarithmic. Also, we centered the histogram bins at logarithmically spaced positions. All permission pairs with a PCP$\,<\!0.001$ have been accounted to the lowest bin.

The plot illustrates that, overall, the PCPs of the simulated applications are significantly lower than those of the real applications.  While the difference for PCPs less or equal than 2\% is small, it grows from this point with increasing PCP. For large PCPs, i.e. with over 50\%, the difference is one order of magnitude large.
Please note that the total number of permission pairs equals for both datasets.
Overall, we compute an average PCP of 8.7\% for the real applications and 1.2\% for the simulated applications. This result suggests that the found patterns  do not randomly arise but might reflect some underlying structure of the Android marketplace.

\section{Discussion}\label{sec_discussion}

We provide a critical discussion of our approach and potential future directions.

{\bf How could this information be conveyed to users?}
Applications that do not fit high-reputation PRPs could be ranked lower in search results, unless they receive enough favorable reviews to override the risk metric.  Another possibility is to alter the permission request UI so that permissions that deviate from PRPs are highlighted, so that the user's attention is focused on the ``unusual'' permission requests.  This would help provide the user with a relative notion of risk.  However, such a design would need to be subject to user research.

{\bf Is this technique suitable for detecting malware?}
We do not aim to provide a  malware detector.  Applications may be untrustworthy, low-quality, buggy, or otherwise risky without being malware.
We found that applications that receive favorable reviews from large numbers of users request permissions differently than applications with few ratings.  Consequently, we designed our approach to identify user satisfaction rather than application maliciousness.

{\bf Why not use supervised learning?}
Our approach uses unsupervised learning to find permission request patterns.
An alternative approach is to manually analyze and label applications based on the characteristics of their permissions or code. In this way, one could create a classifier that discriminates between ``good'' and ``bad'' applications. However, the obvious drawback with this approach is that manually-generated labels are expensive: manual analysis of applications is time-consuming, and there are hundreds of thousands of diverse applications.  Consequently, only a very small set of labels would be available, and the labels might not be representative of all ``good'' or ``bad'' applications. The approach that we propose is fully unsupervised and can automatically learn from the huge corpus of all available applications.

{\bf How indicative for the quality of apps are permissions? } The difference
between high-reputation and low-reputation applications is significant, which we attribute to a difference in application quality. However, other factors could at least partially influence the results.
The low-reputation applications could be newer applications, and permission
request patterns might change over time. This could be accounted for by
computing PRPs for chronologically-similar applications. Some of the
low-reputation applications might become high-reputation applications in the
future; if they were excluded, the difference between the two sets might be more pronounced.  Additionally, low-reputation applications might be highly specialized, leading both to exceptional permission requests and few interested users.  However, we believe that a real difference in applications is the most likely and predominant reason for the disparity in ratings.

{\bf Some permission patterns are requested by only a few applications. How can they be significant?}
Even if only few applications request a pattern, it is still significant if it can be found in different random subsets of applications. We test for this criterion by the stability analysis described in Section~\ref{sec_instab}. In essence, significance means reproducibility or stability and does not necessarily require that a high fraction of applications request the pattern.

\section{Conclusion}

We used a probabilistic model to mine permission request patterns from Android and Facebook applications. For both platforms, we found a set of patterns that fits well to the data.  We found that the permission requests of low-reputation applications differ from the permission request patterns of high-reputation applications. This indicates that permission request patterns can be used as part of a risk metric or soft prediction of the quality of new applications.  For Android, we find that there is a relationship between permission request patterns and categories.  In future work, we will extend the analysis with
application categories in order to achieve greater precision.

{
\section*{Acknowledgment}
This research was supported by Intel through the ISTC for Secure
Computing and by the Swiss National Science Foundation (SNSF), grant no.~138117.

\small

\bibliographystyle{plain}
%\bibliography{biblio}

\begin{thebibliography}{10}

\bibitem{somAndroid}
David Barrera, H.~G \"{u}ne~\c{s} Kayacik, Paul~C. van Oorschot, and Anil
  Somayaji.
\newblock A methodology for empirical analysis of permission-based security
  models and its application to android.
\newblock In {\em Proceedings of the 17th ACM conference on Computer and
  Communications Security}, CCS, 2010.

\bibitem{chia12}
Pern~Hui Chia, Yusuke Yamamoto, and N.~Asokan.
\newblock {Is this App Safe? A Large Scale Study on Application Permissions and
  Risk Signals}.
\newblock In {\em Proceedings of the World Wide Web Conference}, WWW, 2012.

\bibitem{kirin}
William Enck, Machigar Ongtang, and Patrick McDaniel.
\newblock {On Lightweight Mobile Phone Application Certification}.
\newblock In {\em Proceedings of the ACM conference on Computer and
  communications security}, CCS, 2009.

\bibitem{felt-webapps}
Adrienne~Porter Felt, Kate Greenwood, and David Wagner.
\newblock The effectiveness of application permissions.
\newblock In {\em Proceedings of the USENIX Conference on Web Application
  Development}, WebApps, 2011.

\bibitem{Felt:EECS-2012-26}
Adrienne~Porter Felt, Elizabeth Ha, Serge Egelman, Ariel Haney, Erika Chin, and
  David Wagner.
\newblock {Android Permissions: User Attention, Comprehension, and Behavior}.
\newblock Technical Report UCB/EECS-2012-26, EECS Department, University of
  California, Berkeley, 2012.

\bibitem{rbacOrig}
David~F. Ferraiolo and D.~Richard Kuhn.
\newblock {Role Based Access Control}.
\newblock In {\em 15th National Computer Security Conference}, pages 554--563,
  1992.

\bibitem{mtc_ECML}
Mario Frank, Morteza Chehreghani, and Joachim~M. Buhmann.
\newblock The minimum transfer cost principle for model-order selection.
\newblock In {\em ECML PKDD '11}, volume 6911 of {\em Lecture Notes in Computer
  Science}, pages 423--438. Springer Berlin / Heidelberg, 2011.

\bibitem{MAC_jmlr}
Mario Frank, Andreas~P. Streich, David Basin, and Joachim~M. Buhmann.
\newblock Multi-assignment clustering for {B}oolean data.
\newblock {\em Journal of Machine Learning Research}, 13:459--489, Feb 2012.

\bibitem{marketsize}
Leslie Horn.
\newblock Report: Android market reaches 500,000 apps.
\newblock \url{http://www.pcmag.com/article2/0,2817,2395188,00.asp}.

\bibitem{King2011}
Jennifer King, Airi Laminen, and Alex Smolen.
\newblock Privacy: Is there an app for that?
\newblock In {\em Proceedings of the Symposium on Usable Privacy and Security},
  SOUPS, 2011.

\bibitem{Kuhlmann}
Martin Kuhlmann, Dalia Shohat, and Gerhard Schimpf.
\newblock Role mining -- revealing business roles for security administration
  using data mining technology.
\newblock In {\em SACMAT}, pages 179--186, New York, NY, USA, 2003.

\bibitem{Lange2004Stability}
Tilman Lange, Volker Roth, Mikio~L. Braun, and Joachim~M. Buhmann.
\newblock Stability-based validation of clustering solutions.
\newblock {\em Neural Computation}, 16:1299--1323, 2004.

\bibitem{Sanz2012androidappclass}
B.~Sanz, I.~Santos, C.~Laorden, X.~Ugarte-Pedrero, and P.G. Bringas.
\newblock On the automatic categorisation of android applications.
\newblock In {\em Proceedings of the 9$^{th}$ IEEE Consumer Communications and
  Networking Conference (CCNC)}, 2012.

\bibitem{Shabtai:2010:ASC:1931473.1932178}
Asaf Shabtai, Yuval Fledel, and Yuval Elovici.
\newblock Automated static code analysis for classifying android applications
  using machine learning.
\newblock In {\em Proceedings of the 2010 International Conference on
  Computational Intelligence and Security}, CIS, 2010.

\bibitem{socialbakers}
{SocialBakers -- Applications on Facebook}.
\newblock \url{http://www.socialbakers.com/facebook-applications}.

\bibitem{Vaidya:2010:RMP:1805974.1805983}
Jaideep Vaidya, Vijayalakshmi Atluri, and Qi~Guo.
\newblock The role mining problem: A formal perspective.
\newblock {\em ACM Trans. Inf. Syst. Secur.}, 13(3):27:1--27:31, July 2010.

\bibitem{droidranger}
Yajin Zhou, Zhi Wang, Wu~Zhou, and Xuxian Jiang.
\newblock {Hey, You, Get off of My Market: Detecting Malicious Apps in Official
  and Alternative Android Markets}.
\newblock In {\em Proceedings of the Network and Distributed System Security
  Symposium}, NDSS, 2012.

\end{thebibliography}

}

\begin{table}
\centering
\begin{tabular}{| l | }
\hline
PRP 1; requested by 21.46\% of apps; 1.52 bits KL-divergence\\
\hline
Storage : modify/delete USB storage contents \& SD card contents  \\
\hline
\hline
PRP 2; requested by 18.05\% of apps; 1.59 bits KL-divergence\\
\hline
Network communication : full Internet access  \\
Network communication : view network state  \\
\hline
\hline
PRP 3; requested by 15.91\% of apps; 1.49 bits KL-divergence\\
\hline
Network communication : full Internet access  \\
\hline
\hline
PRP 4; requested by 12.41\% of apps; 1.43 bits KL-divergence\\
\hline
System tools : prevent device from sleeping  \\
\hline
\hline
PRP 5; requested by 10.01\% of apps; 1.46 bits KL-divergence\\
\hline
Network communication : full Internet access  \\
Phone calls : read phone state and identity  \\
\hline
\hline
PRP 6; requested by 8.81\% of apps; 1.57 bits KL-divergence\\
\hline
Network communication : full Internet access  \\
Network communication : view network state  \\
Phone calls : read phone state and identity  \\
\hline
\hline
PRP 7; requested by 8.62\% of apps; 1.68 bits KL-divergence\\
\hline
Hardware controls : control vibrator  \\
Network communication : full Internet access  \\
Network communication : view network state  \\
\hline
\hline
PRP 8; requested by 8.55\% of apps; 1.61 bits KL-divergence\\
\hline
System tools : automatically start at boot  \\
\hline
\hline
PRP 9; requested by 8.43\% of apps; 1.42 bits KL-divergence\\
\hline
Network communication : full Internet access  \\
Network communication : view Wi-Fi state  \\
Network communication : view network state  \\
Phone calls : read phone state and identity  \\
\hline
\hline
PRP 10; requested by 8.4\% of apps; 1.57 bits KL-divergence\\
\hline
Network communication : full Internet access  \\
Your location : fine (GPS) location  \\
\hline
\hline
PRP 11; requested by 7.65\% of apps; 1.53 bits KL-divergence\\
\hline
Network communication : full Internet access  \\
Your location : coarse (network-based) location  \\
\hline
\hline
PRP 12; requested by 6.82\% of apps; 1.61 bits KL-divergence\\
\hline
Network communication : full Internet access  \\
System tools : retrieve running applications  \\
\hline
\hline
PRP 13; requested by 6.34\% of apps; 1.4 bits KL-divergence\\
\hline
Hardware controls : control vibrator  \\
\hline
\hline
PRP 14; requested by 5.76\% of apps; 1.32 bits KL-divergence\\
\hline
Network communication : full Internet access  \\
Network communication : view network state  \\
Phone calls : read phone state and identity  \\
Your location : coarse (network-based) location  \\
Your location : fine (GPS) location  \\
\hline
\end{tabular}
\caption{Top requested PRPs for Android.
The frequencies do not sum to 100\% because permission request patterns are not disjoint: permissions can be members of multiple PRPs, and applications can request multiple PRPs.
\label{tab_topPerms}}
\end{table}

\end{document}